\documentclass[review=false,onefignum,onetabnum]{siamonline190516}%

\usepackage{lipsum}
\usepackage{amsfonts}
\usepackage{graphicx}
\usepackage{epstopdf}
\usepackage{algorithmic}
\ifpdf
  \DeclareGraphicsExtensions{.eps,.pdf,.png,.jpg}
\else
  \DeclareGraphicsExtensions{.eps}
\fi

	\usepackage{myMac-SIAM}	

\usepackage{tikz}
	\usetikzlibrary{positioning}
\usepackage{multicol}
\usepackage{multirow}
 		\newcommand{\cheeX}[1]{} 
    	\newcommand{\jingX}[1]{} 
	
		\newcommand{\oldNewX}[2]{#2} 
    \newcommand{\revisedX}[1]{#1}
	\newcommand{\gdisc}{\operatorname{GenDisc}}
    \renewcommand{\wtp}{\breve{p}}	

	\providecommand{\dplus}{D^+}

    \renewcommand{\bsmu}{{\mu}}	
	\newcommand{\bfrho}{{\bs{\rho}}}		
    \newcommand{\ooF}{\mathring{F}}         %
    \newcommand{\e}{e}      %
	
    \newcommand{\ole}{\ol{\e}}      %

\newsiamremark{remark}{Remark}
\newsiamremark{hypothesis}{Hypothesis}
\crefname{hypothesis}{Hypothesis}{Hypotheses}
\newsiamthm{claim}{Claim}

\headers{On $\bfmu$-Symmetric Polynomials}{J.~Yang and C.~Yap}

\title{The D-plus Discriminant and Complexity of Root Clustering\thanks{
	Submitted to arXiv.org on May 9, 2021.
\funding{
	Jing's work is supported by
	National Natural Science Foundation of China (Grant \#11801101)
	and Guangxi Science and Technology Program (Grant \#2017AD23056).
	Chee's work is supported by
	National Science Foundation (Grants CCF-1423228 and CCF-1564132).
	Chee is further supported under Chinese Academy of Science President's
	International Fellowship Initiative (2018),
	and Beihang International Visiting Professor Program No. Z2018060.
}}}

\author{Jing Yang\thanks{
	SMS-KLSE School of Mathematics and Physics,
	Guangxi University for Nationalities, Nanning, China.
  (\email{yangjing0930@gmail.com}).}
\and Chee K. Yap\thanks{
	Courant Institute of Mathematical Sciences, NYU, New York, USA.
  (\email{yap@cs.nyu.edu}).}
  }

\usepackage{amsopn}
\DeclareMathOperator{\diag}{diag}

\ifpdf
\hypersetup{
  pdftitle={The D-plus Discriminant and Complexity of Root Clustering},
  pdfauthor={Jing Yang and Chee Yap}
}
\fi

\begin{document}

\maketitle

\begin{abstract}
	Let $p(x)$ be an integer polynomial
		with $m\ge 2$ distinct roots $\alpha_1\dd \alpha_m$ whose
		multiplicities are $\bfmu=(\mu_1\dd \mu_m)$.
		We define the \dt{D-plus discriminant} of $p(x)$ to be
			$\dplus(p)\as \prod_{1\le i<j\le m}
				(\alpha_i-\alpha_j)^{\mu_i+\mu_j}$.
		Unlike the classical discriminant, $\dplus(p)$
		never vanishes.
	We first prove a conjecture that $\dplus(p)$
		is a $\bfmu$-symmetric function of its roots
		$\alpha_1\dd \alpha_m$.
    Our main result gives an explicit
	formula for $\dplus(p)$, as a rational function
		of its coefficients.
	A basic tool used by our proof is the
	``symbolic Poisson resultant''.
	\revisedX{
	The D-plus discriminant first arose in the complexity
	analysis of a root clustering algorithm from
    Becker et al. (ISSAC 2016).
	The bit-complexity of this algorithm is proportional
	to a quantity $\log(|\dplus(p)|^{-1})$.
	As an application of our main result,
	we give an explicit upper bound
	on this quantity \revisedX{
		in terms of the degree of $p$
		and its leading coefficient.}
	}
\end{abstract}



\sect{Introduction}
\renewcommand{\epf}{\end{pf}{}}

	\cheeX{2Apr'21:
	I rewrote this intro again because the main result
	is very confusing because we mix up $\dplus(p)$ (a number)
	with $\dplus_\bfmu(r_1\dd r_m)$ (a function).  These
	two concepts are now distinct.  Also, I change
	$\dplus(\bfmu)$ to $\dplus_\bfmu$ which is nicer
	when we write ``$\dplus_\bfmu(r_1\dd r_m)$''.

	The new introduction also makes a distinction between
	$\rho_i$'s (roots of $p(x)$) and $r_i$ as variables in a root
	function.  NOTE: this is still NOT perfect because
	later we will use $\rho_i$'s as variables.
	But I think it is OK.}

    \revisedX{
	Consider a polynomial
		$$p=p(x)=\sum_{i=0}^n a_ix^{n-i} \in \ZZ[x],\qquad a_0>0$$
	with $m\ge 1$ distinct
	complex roots $\bfrho=(\rho_1\dd \rho_m)$ and each
	$\rho_i$ has multiplicity $\mu_i\ge 1$.
	Thus $n=\sum_{i=1}^m \mu_i$.
	By reordering the roots, we may assume that
	$\mu_1\ge\mu_2\ge\cdots\ge\mu_m\ge 1$.  Then
	$\bfmu=\bfmu(p)\as (\mu_1\dd \mu_m)$ and
	$\bfrho=\bfrho(p)\as (\rho_1\dd\rho_m)$
	are called the \dt{multiplicity vector}
	and \dt{root vector} of $p$.  We are interested in
	the following \dt{$\bfmu$-discriminant function},
		\beql{dplus}
			\dplus_\bfmu(r_1\dd r_m) \as \prod_{1\le i<j\le m}
			(r_i-r_j)^{\mu_i+\mu_j}.
		\eeql
	The \dt{D-plus discriminant} of $p(x)$ is the
	value obtained by evaluating the $\bfmu(p)$-discriminant function
	at $\bfrho(p)$:
		$$\dplus(p)\as \dplus_\bfmu(\rho_1\dd\rho_m) \in\CC.$$
	We see that $\dplus(p)\in \ol{\ZZ}\ib \CC$
	where $\ol{\ZZ}$ is the algebraic closure of $\ZZ$.
	Although $\dplus_\bfmu$ is defined as a function of the
	roots, the main result in this paper gives an explicit
	polynomial $h_\bfmu(z_1\dd z_n)\in\QQ[z_1\dd z_n]$
	such that
		$$\dplus_\bfmu(r_1\dd r_m) = h_\bfmu(-a_1/a_0\dd (-1)^n a_n/a_0).$$
	Thus proves that $\dplus(p)\in \QQ$.
	Unlike the classical discriminant of a polynomial,
	$\dplus(p)$ cannot vanish.  The non-vanishing property is critical
	in some applications of discriminants.
	For example, Mahler
	\cite{mahler:discriminant:64} shows
	that the \revisedX{minimum} distance between any two roots of $p(x)$
	is proportional to $|D(p)|$ where $D(p)$
	is the classical discriminant of $p$.
	\revisedX{Intuitively, it means that the bit-complexity of
	algorithms to isolate roots of $p$
	is (at least) proportional to $\log(|D(p)|\inv)$.
	}
    \jingX{In the 1964 paper, it was shown that
    the shortest distance between any two roots
    is proportional to $D(p)$ (Theorem 2).
    Is there anything that I missed?
	   CHEE SAYS: sorry, you are right!  Added a
	   sentence to indicate how we move from D(p) to its inverse!}
	The bound becomes trivial when $D(p)=0$.
	For a recent account various
	generalizations of the Mahler bound,
	especially on ways to overcome the case $D(p)=0$,
	see Sharma \cite{sharma:weighted-dmm:20}.

    \begin{xample}
		We illustrate the concept of D-plus discriminant.
	If the polynomial $p(x)=\sum_{i=0}^3 a_{3-i}x^i$ has
	a double root $\rho_1$ and a simple root $\rho_2$, then
	$\dplus(p)=(\rho_1-\rho_2)^3$.  It turns out that
	$\dplus(p)$ is the following rational function in $a_0\dd a_3$:
		\beql{eg}
		\dplus(p)=\efrac{a_0^3}
		\left[a_1^3-(9/2)a_0a_1a_2+(27/2)a_0^2a_3\right].\eeql
	To see a numerical instance,
	let $p(x)=(x-1)^2(x-3)=x^3-5x^2+7x-3$, and so
	$\dplus(p)=(\rho_1-\rho_2)^3=(1-3)^3=-8$.
	Without knowing the roots, \refeq{eg} gives the same answer,
	$\dplus(p)=\left[(-5)^3-(9/2)(-35)+(27/2)(-3)\right]=-8$.
	\end{xample}
	}%

	{\bf History.}
	The root function $\dplus_\bfmu$ was first\footnote{
		In \cite{becker+4:cluster:16}, the D-plus discriminant was
		called a ``generalized discriminant''
		denoted by ``$\gdisc(p)$''.}
	introduced by Becker et al \cite{becker+4:cluster:16}
	in their complexity analysis of a root clustering
	algorithm.
	For simplicity, we may assume\footnote{
		The original algorithm also accepts an input box $B_0\ib\CC$
		specifying a region-of-interest for the roots.
		Our simplification amounts to
		fixing $B_0=[-H,H]^2$ where $H\ge 1+\|p\|_\infty$;
		this ensures that $B_0$ contains all the roots of $p(x)$.
		}
	that their
	algorithm accepts as input the pair $(p,\vareps)$
	where $p\in\CC[x]$
	and $\vareps>0$, and outputs a set $\set{C_i: i=1\dd m}$ where
	$C_i=(\Delta_i,\mu_i)$ represents a cluster of roots.
	Here, the $\Delta_i$'s are pairwise-disjoint
	discs of $\CC$ of radius $<\vareps$, and $\mu_i\ge 1$ is
	the total multiplicity of the roots of $p$ in $\Delta_i$,
	and $n=\sum_{i=1}^m \mu_i$ is the degree of $p$.
	\revisedX{
	To compare the efficiency
	of root isolation algorithms, it is standard to
	use the worst-case bit-complexity $T(n,L)$ for isolating all the
	roots of a polynomial $p$ of degree $n$ with
	$L$-bit integer coefficients: this is
	the so-called ``benchmark problem'' \cite{sagraloff-yap:ceval:11}.
	For the root clustering algorithm in \cite{becker+3:cisolate:18},
	$p$ is not necessarily square-free, and the coefficients may be
	Gaussian integers.  If we
	choose $\vareps$ to be less than $\min{1,\sigma(p)/2}$,
	where $\sigma(p)$ is the minimum separation between distinct
	roots of $p(x)$, we will achieve root isolation,
	i.e., each $\Delta_i$ contains exactly one distinct root.
	}%
	It was shown \cite{becker+3:cisolate:18} that
	that this complexity is ``near-optimal'' in the sense of Pan,
	i.e., $T(n,L)=\wtO(n^2(n+L))$.
	\revisedX{
	Moreover, Imbach-Pan-Yap \cite{imbach-pan-yap:ccluster:18}
	has implemented this algorithm and shown its competitiveness.
	}%
	More precisely, the bit complexity $T(n,L)$ in this case is
         \beql{complexity}
		 \wtO\Big(n^2(L + n + \log(\vareps^{-1}))
			    + n\log|\dplus(p)|^{-1}
			    \Big)
        \eeql
	where $\log(a)\as \max\set{1,\log |a|}$.
	See \cite[Corollary to Theorem A, p.~3, and Appendix A.5]
	{becker+4:cluster:arXiv}.
	Furthermore, the term involving $\dplus(p)$
	can be bounded by $\log|\dplus(p)|^{-1}=O(n(L+\log n))$
	(see Mehlhorn et al \cite[p.~52]{mehlhorn-sagraloff-wang:15}).
	As application of our Main Result,
	we can greatly sharpen this upper bound giving the exact
	dependence on $\bfmu(p)$ and the coefficients of $p(x)$.

	\ignore{
	    \bgenDIY{\sc Corollary to Theorem A}{
	     \ \\The bit complexity of the algorithm is bounded by
	     {
	         \beql{complexity}
	         \wtO\Big(n^2(\tau_F + k+m)
		  		    + nk\LOG(\vareps^{-1})
				    + n\LOG|\gdisc(F_\vareps)|^{-1}
				    \Big).
	        \eeql}
		}

	Hence it is important to give an explicit upper bound
	on this quantity.  As an application of our Main Result,
	we provide such a bound.  Note that Sharma's work
	\cite{sharma:weighted-dmm:20}
	on the ``weighted DMM bounds'' is
	also partly motivated by the clustering
	analysis in \cite{becker+4:cluster:16}.
	}

	\revisedX{
	The key concept in this investigation is the
	concept of a $\bfmu$-symmetric root function, first
	introduced in our companion paper
	\cite{yang-yap:musym:21}.
	We conjectured that
	the root function $\dplus_\bfmu(r_1\dd r_m)$ is
	$\bfmu$-symmetric.
	It follows from this conjecture
	(see Proposition \ref{pro:symmetric} below) that
	$\dplus(p)$ is a rational function in the coefficients of $p(x)$.
	}%
    Using the algorithms developed in \cite{yang-yap:musym:21}
	for detecting $\bfmu$-symmetric polynomials,
	we \revisedX{previously confirmed} the conjecture in many instances.
	In this paper, we will fully prove this conjecture.
    Moreover, the main result of this paper is an explicit
	formula for $\dplus_\bfmu$ as the $(n-m)$-th derivative
	of the classical discriminant of $p$, up to a constant factor
	that depends on $\bfmu$.

\jingX{We need to introduce a new symbol for $K$
    because $K$ is the notation of a field in the rest of the paper.}
\cheeX{Good catch.  I changed $K$ to $C_0$.  But after our zoom,
	we decide to leave it unnamed!}
	\ignore{%
	When $p$ is an integer polynomial, the upper bound on
	$|\dplus(p)|\inv$ is reduced to computing
	a lower bound on $\QQ(\bfmu)$ via a
	recursive optimization argument.
	}%

    The rest of this paper is organized as follows.
	In Section \ref{sec:preliminaries},
    we recall the concepts of $\bfmu$-symmetry and $\bfmu$-gist
    and recall some preliminary properties of $\dplus_\bfmu$.
	In Section \ref{sec:sympoissonres} we develop the tool of symbolic
	Poisson resultant
    for establishing a connection between resultant expressions in roots
    and in coefficients.
    This tool has independent interest.
	In Section \ref{sec:explicitformula},
	we prove our Main Result that $\dplus_\bfmu(r_1\dd r_m)$
	if $\bfmu$-symmetric.  This follows from an
	explicit formula for $\dplus_\bfmu(r_1\dd r_m)$
	in terms of the polynomial coefficients.
    In Section \ref{sec:complexitybound}, we apply the Main Result
	to provide the desired explicit upper bound on
	$\log(|\dplus(p)|\inv)$ in terms of the degree and coefficients
	of $p$.
    We conclude in Section \ref{sec:conclusion}.

\sect{Preliminaries}\label{sec:preliminaries}
	Throughout this paper, we fix $K$ to be
	a domain of characteristic $0$ (in particular, $\ZZ\ib K$).
	Consider three sequences of indeterminates:
		$$\bfx=(x_1\dd x_n),
			\quad \bfz=(z_1\dd z_n),
			\quad \bfr=(r_1\dd r_m)$$
	where $n\ge m\ge 1$.
    Throughout the paper, we use bold fonts for sequences.
	Also we write $|\bfa|$ to denote the length of a sequence $\bfa$.
    In particular, $|\bfmu|=|\bfr|=m$.
	If $S$ is a set, we write
	$\#(S)$ for the number of elements in $S$.

	\revisedX{
	In our application, these indeterminates
	are related through a particular interpretation
	(or specialization)
	as functions of the roots of an arbitrary
	polynomial $p(x)\in K[x]$ of degree $n$ with $m$ distinct roots.
	More precisely, suppose
	$\alpha_1\dd \alpha_n$ are all the roots of $p(x)$
	in $\olK$ (the algebraic closure of $K$).  Then each $x_i$
	specializes to $\alpha_i$, and $z_i$ specializes to
	the $i$th elementary symmetric function of the $\alpha_i$'s.
	Finally, each $r_j$'s specializes to a distinct $\alpha_i$,
	which we also call $\rho_j$.
	}%
    %
\ssect{$\bsmu$-Symmetry and $\bsmu$-Gist}
	Recall the concepts of $\bfmu$-symmetry and $\bfmu$-gist
	from \cite{yang-yap:musym:21}.
	Let $\bfmu=(\mu_1\dd \mu_m)$ where
	$\mu_1\ge \mu_2\ge\cdots\ge\mu_m\ge 1$.
	If $n=\mu_1+\cdots+\mu_m$, we call $\bfmu$ a partition of $n$
	with $m$ parts, or $m$-partition.
	A \dt{specialization} $\sigma$ is any 
	function of the form
		$\sigma:\set{x_1\dd x_n}\to\set{r_1\dd r_m}$.
	We say $\sigma$ is of \dt{type $\bfmu$} if
	$\#(\sigma\inv(r_i))=\mu_i$ for all $i=1\dd m$.
	\revisedX{
	We say $\sigma$ is \dt{canonical} if
	it has the property
	$\sigma(x_i)=r_j$ and $\sigma(x_{i+1})=r_k$
	implies $j\le k$ (for all $i,j,k$).
	For any $\bfmu$, the canonical specialization of type $\bfmu$
	is unique; we denote it by $\sigma_\bfmu$.
	}
	
	Now we consider the polynomial rings $K[\bfx]$ and $K[\bfr]$.
	Any specialization
		$$\sigma:\set{x_1\dd x_n}\to\set{r_1\dd r_m}$$
	can be extended naturally into a $K$-homomorphism
		$$\sigma: K[\bfx]\to K[\bfr]$$
	where $P=P(\bfx)\in K[\bfx]$ is mapped to
	$\sigma(P)= P(\sigma(x_1)\dd \sigma(x_n))$.
	When $\sigma$ is understood, we may write ``$\ol{P}$'' for
	the homomorphic image $\sigma(P)$.

	We denote the \dt{$i$-th elementary symmetric functions}
	($i=1\dd n$) in $K[\bfx]$ by $\e_i=\e_i(\bfx)$.
	For instance, $\e_1 = \sum_{i=1}^n x_i$ and
		$\e_n =\prod_{i=1}^n x_i$.
	Also define $\e_0\as 1$.
    Typically, we write $\ole_i$ for the $\sigma_\bfmu$ specialization
	of $e_i$ when $\bfmu$ is understood from the context;
	thus $\ole_i=\sigma_{\bfmu}(\e_i)\in K[\bfr]$.

	\begin{defn}
        A polynomial $F\in K[\bfr]$ is said to be \dt{$\bfmu$-symmetric}
        if there exists a symmetric polynomial $\whF\in K[\bfx]$ such that
        $\sigma_\bfmu(\whF)=F$.
	\end{defn}

	\revisedX{
	When $F$ is $\bfmu$-symmetric, we can define a related
	polynomial as follows: since $\whF\in K[\bfx]$ is symmetric,
	the Fundamental Theorem of Symmetric Functions tells
	us that there exists a polynomial
        $\ooF\in K[\bfz]$ such that
		$\ooF(\e_1\dd \e_n)=\whF(\bfx)$.
		We call $\ooF$ the \dt{$\bfmu$-gist} of $F$.
		Moreover,
			\beql{gistF}
				\ooF(\ole_1\dd \ole_n)=F(\bfr)\eeql
		where $\ole_i=\sigma_\bfmu(e_i)$.
	}
	\revisedX{
	E.g., let $\bfx=(x_1\dd x_4),\bfr=(r_1,r_2,r_3)$,
	$\bfz=(z_1\dd z_4)$ and $\bfmu=(2,1,1)$.
	Consider $F(\bfr)=(2r_1+r_2+r_3)^2+r_1^2r_2r_3$. It
	is $\bfmu$-symmetric since there is a symmetric polynomial
	$\whF(\bfx)=(\sum_{i=1}^4 x_i)^2+\prod_{i=1}^4 x_i=e_1^2+e_4$
	such that $F(\bfr)=\ole_1^2+\ole_4=\sigma_\bfmu(\whF(\bfx))$.
	The $\bfmu$-gist of $F(\bfr)$ is $\ooF(\bfz)=z_1^2+z_4$
	since $F(\bfr)=\ooF(\ole_1\dd \ole_4)$.
	}

\cheeX{ New on 8Mar'21: The conclusion of this proposition
	is what we were after, but it was never
	explicitly stated anywhere!  Now, the introduction
	cites this.  We must also cite this proposition
	when discussing $\dplus$ below.}
	\revisedX{
	The following proposition from
	\cite[Proposition 1]{yang-yap:musym:21}
	shows that if $F$ is
	$\mu(p)$-symmetric function of the distinct roots of $p(x)$, then
	$F$ is a rational function in the coefficients of $p(x)$:
	}
	\bpro
    {
		\label{pro:symmetric}
		Let $F(\bfr)\in K[\bfr]$ be $\bfmu$-symmetric with
		gist $\ooF(\bfz)\in K[\bfz]$.
		For any polynomial
			$$p(x)=\sum_{i=0}^n a_i x^{n-i}\in K[x]$$
	with $m$ distinct roots $\bfalpha=(\alpha_1\dd \alpha_m)$
	of multiplicity $\bfmu$,
		$$F(\bfalpha)=\ooF(-a_1/a_0, a_2/a_0\dd (-1)^na_n/a_0)$$
	Thus, $F(\bfalpha)$ is a polynomial in $(a_1/a_0\dd a_n/a_0)$.
	}\epro
	\ignore{
	\bpf
		Let $\alpha_{m+1}\dd \alpha_n$ be the remaining roots of $p(x)$
	and let $\sigma:\set{1\dd n}\to\set{1\dd m}$ such
	that $\#\sigma\inv(i)=\mu_i$ ($i=1\dd m$).
	If $e_j=e_j(\alpha_1\dd\alpha_n)$ is the $j$th elementary symmetric
	function, then $e_j=\ole_j$ where $\ole_j=\ole_j(\bfalpha)$ is
	a function of the distinct roots:
		$\ole_i(\bfalpha)
			\as e_i(\alpha_{\sigma(1)}\dd\alpha_{\sigma(n)})$.
	Then $p(x)=c_0\sum_{i=1}^n(x-\alpha_i)
		= c_0\sum_{j=0}^n(-1)^j\ole_j x^{n-j}$.
	Equating coefficients in these two expressions for $p(x)$,
	we conclude that $c_j = c_0(-1)^j \ole_j$.
	By definition of the gist of $F$,
		\beql{fbfalpha}
			F(\bfalpha) = \ooF(\ole_1\dd \ole_n).\eeql
	Our Proposition follows when we plug $(-1)^ic_i/c_0$ for $\ole_i$
	in \refeq{fbfalpha}.
	\epf
	}
\ssect{Basic Propositions on the D-plus function}
	Recall the definition of $\dplus_\bfmu=\dplus_\bfmu(\bfr)$
	in \refeq{dplus}.
    We say $\dplus_\bfmu$ is \emph{injective}
	if for all $\bfmu'\neq\bfmu$, $\dplus_\bfmu\neq \dplus_{\bfmu'}$.
    Clearly, $\dplus_\bfmu$ is not injective if
    $\bfmu=(\mu_1,\mu_2)$ and $n=\mu_1+\mu_2\ge 3$:
    in this case, $\dplus=(r_1-r_2)^n$ for any $\bfmu=(\mu_1,\mu_2)$
    where $\mu_1+\mu_2=n$.
    The next proposition shows that $\dplus_\bfmu$ is injective
	in all other cases:

    \bpro
    $\dplus_\bfmu$ is injective whenever $|\bfmu|\ge 3$.
    \epro
    \bpf
	Let $\bfmu=(\mu_1\dd\mu_m)$ where $m \ge 3$.
    Suppose $\bfmu'=(\mu'_1\dd \mu'_{m'})$ such that
	$\dplus_\bfmu=\dplus_{\bfmu'}$.
	We must prove that $\bfmu=\bfmu'$.
	By definition of $\dplus$, it is clear that $m'$ is equal to $m$.
	Define $\delta_i \as \mu_i'-\mu_i$.
	The equality $\dplus_\bfmu=\dplus_{\bfmu'}$ implies that
    $\mu_i+\mu_j=\mu_i'+\mu_j'$ for all $i\neq j$.
	It follows $\delta_i=-\delta_j$ for all $i,j$.
	Since $m\ge 3$, we conclude that
		\beqarrays
			\mu_2'+\mu_3' &=& (\mu_2+\delta_2)+(\mu_3+\delta_3)\\
					&=& (\mu_2-\delta_1)+(\mu_3-\delta_1)\\
					&=& (\mu_2+\mu_3)-2 \delta_1
		\eeqarrays
	which implies that $\delta_1=0$.  Hence $\delta_i=-\delta_1=0$
	for all $i$.  This is equivalent to $\bfmu=\bfmu'$.
    \epf

	In \cite{yang-yap:musym:21}, we proved the $\bfmu$-symmetry
	of $\dplus_\bfmu$ for the following two special cases.

    \bpro
    Assume $\bfmu=(\mu_1,\mu_2)$ satisfying $\mu_1+\mu_2=n$.
    Then $\dplus_\bfmu$ is $\bfmu$-symmetric with gist
	    \[
			\mathring{\dplus_\bfmu}(\bfz) =\left\{
	    \begin{array}{ll}
	    \Big( \frac{(n-1)z_1^2-2n z_2}{
					\mu_1\mu_2}\Big)^{n/2}
	    & \rmif\ n\mbox{~is even,}\\[15pt]
	    \Big( \frac{(n-1) z_1^2-2n z_2}{
					\mu_1\mu_2}\Big)^{\frac{n-3}{2}}
				\Big( k_1 z_1^3
					+ k_2 z_1 z_2+k_3 z_3\Big)
	    & \rmif\ n\mbox{~is odd.}
	    \end{array}
	    \right.
	    \]
    where $k_1= \frac{-(n-1)(n-2)}{d},
			k_2= \frac{3n(n-2)}{d},
			k_3= \frac{-3n^2}{d}$
			and $d=\mu_1\mu_2(\mu_1-\mu_2)$.
	\epro

    \bpro
    If all $\mu_i$'s are equal to $\mu$, then
    $\dplus(P(x))$ is $\bfmu$-symmetric with gist given by
		\[
			\mathring{\dplus_\bfmu}(\bfz)
				=\left(\frac{1}{\mu^{m}}\cdot S^n_{n-m}\right)^\mu
		\]
    where $S^n_{n-m}$ is the $(n-m)$th subdiscriminant of $P$
    of degree $n$.
    \epro	
\sect{The Symbolic Poisson Resultant}\label{sec:sympoissonres}
	Our proof of the $\bfmu$-symmetry of $\dplus$ involves two technical
	tools: differentiation in a quotient ring and
	the symbolic version of the Poisson resultant:
	\benum[(I)]
    \item
		\revisedX{
	Let $R$ be a ring and $c$ an indeterminate.
	We treat the polynomial $R[c]$
	as a differential ring with a differential
	operator $\del[c]: R[c]\to R[c]$ where $\del[c](c)=1$
	and $\del[c](a)=0$ ($a\in R$).  We will need to extend
	$\del[c]$ to the quotient ring $R[c]/I$ where
	$I\ib R[c]$ is an ideal.
	Let the equivalence class of $a\in R[c]$ modulo $I$
	be denoted as $[a]\in R[c]/I$.  Let us define $\del[c]([a])$ to be
	$[\del[c](a)]$.  It is easy to check that $\del[c]([a])$
	is well-defined iff $I$ is closed under $\del[c]$.
	}%
    \item
		\revisedX{
	The resultant $\res(A,B)$ of polynomials $A,B\in K[x]$
	has an algebraic-geometric dual nature in the following sense:
	as an algebraic object, $\res(A,B)$ is defined as an expression
	in terms of the coefficients of $A$ and $B$.
	Hence $\res(A,B)\in K$.
	As a geometric object, it is defined as an expression
	$\ol{\res}(A,B)$ in terms of
	$\alpha_i$'s and $\beta_j$'s which are the roots of
	$A$ and $B$ in $\olK$ (the algebraic closure of $K$).
	Hence $\ol{\res}(A,B)\in\olK$.
	We may call $\ol{\res}(A,B)$ the \dt{Poisson resultant}.
	Naturally, the algebraic and geometric concepts coincide:
	$\ol{\res}(A,B)=\res(A,B)$.  In reference to this equality,
	we may call
	$\ol{\res}(A,B)$ the \dt{Poisson formula} for $\res(A,B)$.
	See \cite[Chap.~VI.4]{yap:algebra:bk} for this classical approach.
	In this paper, in order to give a rigorous proof
	of our $\dplus$-theorem (\refThm{dplusismusymmetric} below), we
	introduce a symbolic version of the Poisson formula in which
	the $\alpha_i$'s and $\beta_j$'s are just symbols.
	}%
	\eenum
	
	We now state and prove the symbolic Poisson formula.
	Assume $n,m\ge 1$.
	Let $A=\sum_{i=0}^m a_i x^{m-i}$,
	and $B=\sum_{j=0}^n b_j x^{n-j}$
	where $x$,
	$\bfa=(a_0\dd a_m)$ and $\bfb=(b_0\dd b_n)$ are $m+n+3$
	indeterminates.
	\revisedX{
	We define $\res(A,B)=\res_x(A,B)$ (eliminating $x$)
	as the determinant of the Sylvester matrix $M(A,B)$
	\cite[Chap.~III.3]{yap:algebra:bk}.
	The non-zero entries of $M(A,B)$ contain the coefficients
	of $A,B$.  Therefore $\res(A,B)\in \ZZ[\bfa,\bfb]$.
	}%

	Next we introduce two sets of indeterminates,
		$\bfalpha=(\alpha_1\dd \alpha_m),\quad
		\bfbeta=(\beta_1\dd \beta_n)$
	and the corresponding Viete relations expressing
	them as roots of $A$ and $B$:
		\beqarrys
		V^a(\bfa,\bfalpha) &\as&
			\set{ a_i - (-1)^i e_i(\bfalpha)a_0:i=1\dd m},\\
		V^b(\bfb,\bfbeta) &\as&
			\set{ b_j - (-1)^j e_j(\bfbeta)b_0:j=1\dd n}
		\eeqarrys
	where $e_i(x_1\dd x_k)$ is the $i$-th elementary symmetric
	function in $k$ variables; note that
	$k=m$ in $V^a$, and $k=n$ in $V^b$.
	\revisedX{
	For any ring $R$ and subset $S\ib R$, let
	$\bang{S}=\bang{S}_R$ denote the ideal generated by $S$ in $R$.
	If $p,q\in R$,
	we say \dt{$p$ is equivalent to $q$ modulo $S$}
	(written as ``$p\equiv q$ (mod $S$)'') if $p-q\in \bang{S}$.
	}%
	We are ready to state the symbolic
	Poisson formula for the resultant:

	\bthmT{Symbolic Poisson Formula}{poisson} \ \\
		\revisedX{
	Let $k$ be a symbol in $\set{a,b,ab}$.
	Define $Q_k$, $R_k$ and $J_k$ as in the following table:
		\btable[llllll]{
		             &Polynomial & Ring & Ideal \\\hline
      	$\mathrm{(a)}$ & $Q_a\as a_0^n\prod_{i=1}^m B(\alpha_i)$
					 & $R_a\as \ZZ[\bfa,\bfb,\bfalpha]$
			         & $J_a\as \bang{V^a}$ \\
	 $\mathrm{(b)}$ &$Q_b\as (-1)^{mn}b_0^m\prod_{j=1}^n A(\beta_j)$
					 & $R_b\as \ZZ[\bfa,\bfb,\bfbeta]$
			         & $J_b\as \bang{V^b}$ \\
	$\mathrm{(c)}$ &$Q_{ab}\as a_0^nb_0^m\prod_{i=1}^m\prod_{j=1}^n
			(\alpha_i-\beta_j)$
					 & $R_{ab}\as \ZZ[\bfa,\bfb,\bfalpha,\bfbeta]$
					 & $J_{ab}\as \bang{V^a\cup V^b}$
		}
	Then the resultant polynomial $\res(A,B)\in\ZZ[\bfa,\bfb]$
	is equivalent to $Q_k$ modulo $J_k$ in the ring $R_k$.
		}
	\ethmT
	\revisedX{
	\bpf
	Note that $\res(A,B)$ and $Q_k$ are polynomials
	in the ring $R_k$, and $J_k$ an ideal of $R_k$.
	The theorem makes three claims: for each $k\in\set{a,b,ab}$,
	we claim that $\res(A,B)-Q_k\in J_k$ in $R_k$.
	Two types of arguments will be used in the proof:
	Type (I) depends on the Viete relations, and
	Type (II) is based on algebraic identities on determinants
	and the resultant polynomial.
	Type (II) justifications are summarized as follows
	(see \cite[Lemmas~VI.6.12, VI.6.14]{yap:algebra:bk}):
		\benum[\qquad\qquad{II}(a):]
		\item $\res(\alpha,B)=\alpha^{n}$.
		\item $\res(A,B) = (-1)^{mn}\res(B,A)$.
		\item $\res(\alpha A,B) = \alpha^{n}\res(A,B)$.
		\item $\res((x-\alpha)A,B) = B(\alpha)\cdot\res(A,B).$
		\eenum
	In the following proof, we write ``$\equiv$'' for a Type (I)
	justification, and `$=$' for Type (II) justification:
	
	\begin{align*}
		\res(A,B) & \equiv \res(a_0\prod_{i=1}^m (x-\alpha_i), B)
						& \text{(modulo $\bang{V_a}$)}\\
				& = a_0^n\cdot\res(\prod_{i=1}^m(x-\alpha_i), B)
						& \text{(by II(c))}\\
				& = a_0^nB(\alpha_1)\cdot\res(\prod_{i=2}^m(x-\alpha_i), B)
						& \text{(by II(d))}\\
				& ~~~\vdots &\\
				& = a_0^n\prod_{i=1}^m B(\alpha_i)\cdot\res(1, B)
						& \text{(by II(d))}
    \end{align*}
	\begin{align*}
				& = a_0^n\prod_{i=1}^m B(\alpha_i)
						& \text{(by II(a))}
	\end{align*}
	This proves the claim when $k=a$;
	the proofs for $k=b$ and $k=ab$ are similar.	
	\epf
	}

\begin{figure}[h]\caption{The commutative diagram of
		$R_i$ and $R_{ij}$}\label{fig:diagram}
    	\begin{center}
			\tikzset{node distance=9mm, auto}
		\begin{tikzpicture}
		\node (r0) {$R_0$};
		\node (r1) [below left=of r0] {$R_1$};
		\node (r2) [below right=of r0] {$R_2$};
		\node (r12) [below =of r1] {$R_{12}$};
		\node (r21) [below =of r2] {$R_{21}$};
		\node (r3) [below right =of r12] {$R_{3}$};
		\node (J1J2) [below =15mm of r0] { $J_1+J_2$ };
		\draw[->] (r0) to node[swap] { $J_1$ } (r1);
		\draw[->] (r0) to node { $J_2$ } (r2);
		\draw[->] (r1) to node[swap] { $[J_2]_1$ } (r12);
		\draw[->] (r2) to node { $[J_1]_2$ } (r21);
		\draw[-] (r0) to node {} (J1J2);
		\draw[->] (J1J2) to node {} (r3);
		\draw (r12) to node[swap] { $\simeq$ } (r3);
		\draw (r21) to node { $\simeq$ } (r3);
		\end{tikzpicture}
		\end{center}
    \end{figure}
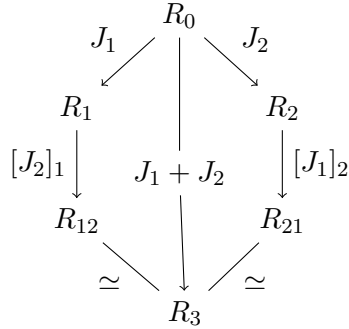

	\cheeX{I streamlined the next lemma, and gave it a more
		"geometric" (less "algebraic") proof.
		Also tried to explain why we need this lemma.}
	\revisedX{
	In the previous theorem, we proved three equivalences:
	that $\res(A,B)$ is equivalent (respectively)
	to $Q_a$ in $R_1\as R_{ab}/\bang{V_a}/[\bang{V_b}]$,
	to $Q_b$ in $R_2\as R_{ab}/\bang{V_b}/[\bang{V_a}]$, and
	to $Q_{ab}$ in $R_3\as R_{ab}/\bang{V_a\cup V_b}$.
	In the non-symbolic world, these three separate equivalents
	are simply identities over a common ring $\olD$.
	To achieve a similar ``flattening'' in the symbolic
	setting, we use the fact that the three rings $R_1, R_2, R_3$ are
	isomorphic.  This follows from the following lemma:
	\bleml{iso}
	Consider a ring $R_0$ with ideals $J_i$ ($i=1,2$).
	Let $R_i\as R_0/J_i$ with the induced ideal
		$J_{ij} = \set{b+J_i: b\in J_j} \ib R_i$ ($i\neq j$).
	Then the following three rings are isomorphic:
		$$R_1/J_{12} \simeq R_0/(J_1+J_2) \simeq R_2/J_{21}$$
	See Figure \ref{fig:diagram}, where we denote
	$R_i/J_{ij}$ by $R_{ij}$ and $R_0/(J_1+J_2)$ by $R_3$.
	\eleml
	}%

	\revisedX{%
	\bpf
	By symmetry, it is enough to show that
	$R_1/J_{12}$ and $R_0/(J_1+J_2)$ are isomorphic.
    For any $a\in R_0$, define the cosets
		$$\begin{array}{lllll}
			~[a]	& \as~ a+(J_1+J_2) 	& \in R_0/(J_1+ J_2),\\
			~[a]_1	& \as~ a+J_1			& \in R_1,\\
			~[a]_2	& \as~ [a]_1+J_{12}	& \in R_1/J_{12}.
		\end{array}$$
	Note that $[a]$ and $[a]_1$ are subsets of $R_0$, but $[a]_2$
	is a set of subsets of $R_0$.
	If we flatten $[a]_2$ into $\ol{[a]_2}\as \cup_{b\in J_2} [b]_1$,
	we have the equality $\ol{[a]_2}=[a]$.
	This gives a bijection between $R_0/(J_1+J_2)$ and $R_1/J_{12}$.
	In other words, the maps $a\mapsto [a]$ and $a\mapsto [a]_2$
	are homomorphisms from $R_0$ to $R_0/(J_1+J_2)$ and $R_1/J_{12}$:
	we have $[a+a']=[a]+[a']$ and $[aa']=[a][a']$
	and similarly for $[a+a']_2$ and $[aa']_2$.
	This proves the isomorphism $R_1/J_{12}\simeq R_0/(J_1+J_2)$.
    \epf
	}%

	\dt{Remark:}
	Figure \ref{fig:diagram} shows two pathways
	from $R_0$ to an isomorphic copy of $R_3$:
	either via $R_0/J_1$ or via $R_0/J_2$.
	In the next section, our proof allows only one
	of the two pathways.
\sect{Main Result: A Gist Formula for $\dplus_\bfmu$}
\label{sec:explicitformula}
	\cheeX{Feb'21:
	I have greatly re-arranged the content of this section
	because previously, the notations were introduced piecemeal,
	and it hard to track.  We now put it all upfront!}
	\cheeX{29Mar'21:
	I found the main theorem unsatisfactory for the following
	reason: we talked about $D^+(p)$ where
		$p(x)=\sum_{i=0}^n c_ix^{n-i}\in K[x]$.
		Thus the coefficients of $p(x)$ are numbers in $K$.
	But we need to differentiate $D=\res_x(p,p')$
	with respect to $c_n$, which is a number, not a variable.
	Solution: we need a "symbolic" version of
	$p(x)$.   So we now begin with
		$p(x)=\sum_{i=0}^n a_ix^{n-i}\in K[x]$,
		but introduce
	$\wtp(x)\as \sum_{i=0}^n c_ix^{n-i}\in K[\bfc][x]$
	with symbolic coefficients $c_i$.
	Moreover, the main result says that
	$\bfmu$-discrimininant is $\bfmu$-symmetric
	and gives the gist formula in terms of the coefficients $c_i$'s.
	}
	\cheeX{2May'21:
	Again, I find it easier to understand if
	we explicitly add $\bfr=(r_1\dd r_m)$
	to the definition of the underlying ring $R_0$.
	The introduction has been re-written to clearly
	distinguish between the ``D-plus discriminant
	root function $\dplus_\bfmu(\bfr)$''
	and the complex number
	``D-plus discriminant $\dplus(p)\in \CC$''
	of $p(x)\in \ZZ[x]$.
	}

	\revisedX{
    Fix a multiplicity vector
	$\bfmu=(\mu_1\dd \mu_m)$,
	and the variables $\bfr=(r_1\dd r_m)$
	and $\bfz=(z_1\dd z_n)$ where $n=\sum_{i=1}^m \mu_i$.

	In this section, we prove the main result of this paper:
	the $\bfmu$-discriminant function $\dplus_\bfmu(\bfr)$
	is a $\bfmu$-symmetric function.
	By \refPro{symmetric}, $\dplus_\bfmu(\bfr)$
	has a $\bfmu$-gist function
	$\mathring{\dplus_\bfmu}(\bfz)\in \ZZ[\bfz]$.
	This means that if $p(x)=\sum_{i=0}^n a_ix^{n-i}\in K[x]$
	has $\bfmu$ as multiplicity vector, then
		$$\dplus(p)=
			\mathring{\dplus_\bfmu}(-a_1/a_0\dd (-1)^na_n/a_0).$$
	We further provide an explicit formula for
	$\mathring{\dplus_\bfmu}(\bfz)$.  To do this, we
	need to introduce the symbolic counterpart of $p(x)$,
		\beql{wtp}
			\wtp(x)\as \sum_{i=0}^n c_ix^{n-i}\in K[\bfc][x]
			\eeql
	where $\bfc=(c_0,c_1\dd c_n)$ are indeterminates.
	Then we take the discriminant of $\wtp$
	(\cite[Chapter VI.6]{yap:algebra:bk}):
		\beql{D}
			D(\wtp)\as \frac{(-1)^{n\choose 2}}{c_0}\res_x(\wtp,\wtp')
			\eeql
	where $\wtp'$ denotes the differentiation of $\wtp$ with respect to
	$x$.  Finally, we need to perform two types
	of operations on $D(\wtp)$, differentiation and specialization.
	These operations on $F\in K[\bfc][x]$ are defined as follows:
	\bitem
	\item
		Let $\del[n] F$ denote the partial derivative of $F\in K[\bfc][x]$
	with respect to $c_n$.  For $k\ge 0$, let
		$$\del[n]^k F =\clauses{ F &\rmif\ k=0,\\
							\del[n](\del[n]^{k-1} F) &\rmif\ k\ge 1.}$$
	\item
		Consider the specialization
			\beql{c_i}
				\set{c_i\to (-1)^i z_i c_0 : i=1\dd n}\eeql
		which introduces the variables $\bfz=(z_1\dd z_n)$.
		It induces a homomorphism from $K[\bfc][x]$ to $K[c_0,\bfz][x]$
		where $F\in K[\bfc][x]$ specializes to
			$$F\big|_{c_i\ass(-1)^iz_ic_0~(i=1\dd n)}\in
				K[c_0,\bfz][x].$$
	\eitem

	We are ready to state our result.

	\bthmT{Main Result}{dplusismusymmetric}
		\ \\
		The $\bfmu$-discriminant function
		$\dplus_\bfmu(\bfr)$ is $\bfmu$-symmetric, and has
		a $\bfmu$-gist given by
		\beql{H}
			\mathring{\dplus_\bfmu}(\bfz)=\efrac{C_{\bfmu}}H(\bfz)
			\eeql
	where
		\begin{align}
			C_{\bfmu}& \as (n-m)!
				(-1)^{mn+{n\choose 2}+\sum_{i=1}^m{i\mu_i}}
				\cdot \prod_{i=1}^m\mu_i^{\mu_i},
					\label{eq:Cbfmu}\\
			H(\bfz)=H_{n,m}(\bfz)& \as \frac{1}{c_0^{m+n-2}}\cdot
				\big(\del[n]^{n-m}D(\wtp)\big)
				\Big|_{c_i\ass(-1)^iz_ic_0~(i=1\dd n)}.
					\label{eq:Hbfz}
		\end{align}
    \ethmT
	}

	Note that the constant $C_\bfmu$ is an integer,
	and the polynomial $H(\bfz)$ depends only on $n$ and $m$,
	and does not depend on the $\mu_i$ values.
    Before giving the proof of this theorem,
	we illustrate the gist formula of $\dplus_\bfmu$:

    \begin{xample}
		Continue with $p(x)=\sum_{i=0}^3 a_{3-i}x^i\in K[x]$ from
		Example 1 with multiplicity vector $\bfmu=(2,1)$.
		Thus $n=3, m=2$.
    \bitem
    \item
		The symbolic counterpart of $p(x)$ is
		$\wtp(x)=\sum_{i=0}^3 c_{3-i}x^i \in K[\bfc][x]$.
    \item
	\oldNewX{$c~~$} {$C_\bfmu$}
    $=(-1)^{mn+\frac{n(n-1)}{2}+\sum_{i=1}^m{i\mu_i}}
        \cdot (n-m)!\prod_{i=1}^m\mu_i^{\mu_i}=-4$
    \item \revisedX{
		$D(\wtp)=\frac{(-1)^{3\choose 2}}{c_0}\res_x(\wtp,\wtp')
            =-4\,{c_{{1}}^{3}}c_{{3}}
                +{c_{{1}}^{2}}{c_{{2}}^{2}}
                +18\,c_{{0}}c_{{1}}c_{{2}}c_{{3}}
                -4\,c_{{0}}{c_{{2}}^{3}}
                -27\,{c_{{0}}^{2}}{c_{{3}}^{2}}$}
    \item \revisedX{
			$\del[3]^{3-2}D(\wtp)=\del[3]D(\wtp)= -4\,{c_{{1}}^{3}}
                        +18\,c_{{0}}c_{{1}}c_{{2}}
                        -54\,{c_{{0}}^{2}}c_{{3}}$}
    \eitem
    Then
		\revisedX{
		$$H=\frac{(-4\,{c_{{1}}^{3}}
            +18\,c_{{0}}c_{{1}}c_{{2}}
            -54\,{c_{{0}}^{2}}c_{{3}})|_{c_i=(-1)^iz_ic_0}}{
			{c_0^{2+3-2}}}=4z_1^3-18z_1z_2+54z_3.$$
			}
	The substitution of $z_i=\ole_i$ into $H$ yields
    \begin{align*}
    H(\ole_1,\ole_2,\ole_3)
    &=4(2r_1+r_2)^3-18(2r_1+r_2)(r_1^2+2r_1r_2)+54r_1^2r_2\\
    &=-4(r_1-r_2)^3=C_\bfmu\dplus.
    \end{align*}
	Note that this derivation agrees with Example 1~in \refeq{eg}.
\cheeX{ Why is it "easy to check''?  Perhaps give
	a brief intermediate calculation?  }
    \end{xample}

	\revisedX{
	\dt{Setup for proof of the Main Result}:
	\benum[(S1)]
	\item
		Our entire proof takes place inside the
		polynomial ring $R_0[x]$ where
		$$R_0\as\ZZ[\bfc,\bfalpha,\bfbeta,\bfr]$$
	where
	$\bfc\as (c_0\dd c_n)$,
	$\bfalpha\as (\alpha_1\dd \alpha_n)$,
	$\bfbeta\as (\beta_1\dd \beta_{n-1})$ and $\bfr\as (r_1\dd r_m)$
	are indeterminates.  Call $R_0$ the \dt{underlying domain}.
	\item
	We will introduce three quotient rings
	$R_1, R_2, R_3$ which are connected via canonical homomorphisms,
		\beql{hi}
			h_i: R_{i-1}\to R_i,\qquad (i=1\dd 3).
		\eeql
	Also, let $h_0:R_0\to R_0$ be the identity.
	We unify these homomorphisms by focusing on the underlying domain
	$R_0$, and define $\olh_k:R_0\to R_k$ ($k=0,1,2,3$) where
			$$\olh_k=\clauses{ h_0 & \rmif\ k=0,\\
						h_k\circ \olh_{k-1} & \rmif\ k\ge 1}$$
	where $\circ$ denotes function composition.
	Let $a,b\in R_0$.
	We usually prefer to write $[a]_k$ instead of $\olh_k(a)$.
	We say $a,b$ are \dt{$\olh_k$-equivalent}, denoted as
	$a\equiv_k b$, if $[a]_k=[b]_k$.
	Clearly,
		$a\equiv_k b$ implies $a\equiv_{k+1} b$ for $k=0,1,2$.
	If $S\ib R_0$, we write $\bang{S}_k$ for the ideal in $R_k$
	generated by $\set{[a]_k: a\in S}$.
	\item
		Definition of $R_1$ and $R_2$:
	we establish $\bfalpha$ (resp., $\bfbeta$) as the
	roots of $\wtp$
	(resp., $\wtp'(x)=\sum_{i=0}^{n-1} (n-i)c_i x^{n-1-i}$)
	using the Viete relations:
		\beql{v}
		\clauses{
			V &\as \{c_i-c_0\cdot(-1)^ie_i(\bfalpha):\,i=1\dd n\},\\
			V'&\as\set{(n-i)c_i -(nc_0)(-1)^i e'_i(\bfbeta): i=1\dd n-1}.}
		\eeql
	\oldNewX{(error in using $e_j$)
		}{where $e_i$ (resp.,~$e'_i$) is the $i$th elementary symmetric
		function on $n$ (resp.,~$n-1$) variables}.
	This gives us two ideals $\bang{V}_0, \bang{V'}_0 \ib R_0$,
	as in \refLem{iso} (see Figure \ref{fig:diagram}).
	Although we may take successive quotients of $R_0$
	by $V$ and $V'$ in either order, our proof requires that we first
	take quotient by $V'$, as follows:
		\beql{r1r2} R_1\as R_0/\bang{V'}_0,
			\qquad R_2\as R_1/\bang{V}_1	\eeql
	\item
		Definition of $R_3$:
	Finally, we define
		\beql{r3}
			R_3\as R_2/\bang{W\cup W'}_2,
			\eeql
	where $W$ are relations asserting that $\bfalpha$
	has $m$ distinct (symbolic) roots $\bfr=(r_1\dd r_m)$,
	and $W'$ are relations connecting the roots $\bfbeta$ to
	$\bfalpha$.
	More precisely, let
		\beql{sigma}
			\sigma_\bfmu:
			\set{\alpha_{1}\dd\alpha_n}\to\set{r_1\dd r_m}\eeql
	be any specialization that maps exactly
	$\mu_j$ of the $\alpha_i$'s to each $r_j$ ($j=1\dd m$).
	Then
		\beql{w}
			W \as \set{\alpha_i - \sigma_\bfmu(\alpha_i): i=1\dd n}.
		\eeql
	Next consider the relations that
	determine the multiplicity of those $\beta_j$'s that
	are equal to one of $r_i$'s.  We know that
	$n-m = \sum_{i=1}^m (\mu_i-1)$ of the $\beta_j$'s
	are mapped to $r_1\dd r_m$.  The remaining
	$m-1$ $\beta_j$'s are not equated with any root of $p$; wlog,
	let\footnote{
		Of course, it does not mean that $\beta_1\dd \beta_{m-1}$
		are unrelated to $r_1\dd r_m$.  But the connection
		is more subtle, and this is captured by Lemma C below.
		}
	$\beta_1\dd \beta_{m-1}$
	be these nonroots of $p$.  These relations are
	captured by the specialization
		\beql{sigma'}
			\sigma_\bfmu':\set{\beta_m\dd \beta_{n-1}}\to
			\set{r_1\dd r_m}\eeql
	map exactly $\mu_j-1$ of the $\beta_i$'s in $\bfbeta\backslash\bfbeta'$ to
	each $r_j$ ($j=1\dd m$).  Thus our desired set of
	relations $W'$ is given by
		\beql{w'}
			W' = \set{\beta_i - \sigma_\bfmu'(\beta_i): i=m\dd n-1}.
		\eeql
	\eenum
	}

\cheeX{REVISIONS IS STABLE UP TO THIS POINT:}

	\revisedX{
	\dt{Proof of the Main Result:}
	The arguments will be derived
	through a sequence of four lemmas (A,B,C and D).
	In Lemma A, we will show that
	the $(n-m)$th derivative of the discriminant of $\wtp$,
		\beql{g} G\as \del[n]^{n-m}D(\wtp)
			\eeql
	can be expressed in terms of the roots $\bfalpha,\bfbeta$
	by an application of the symbolic Poisson formula:
	}
	
	\blemDIY[Lemma A]{
					\label{gequiv} 
	With $G$ as defined in \refeq{g}, we have
		\beql{g2}
		G\equiv_2 (n-m)!C\cdot  \sum_{S\in{[n-1]\choose m-1}}
				\prod_{j\in S} p_\bfalpha(\beta_j)\eeql
    where
		${[n-1]\choose m-1}$ denotes the set of
		$(m-1)$-sets contained in $[n-1]\as\set{1\dd n-1}$, and
	\begin{align}
		C	& \as (-1)^{\frac{n(n-1)}{2}}n^nc_0^{n-1}, \label{eq:C}\\
		p_{\bfalpha}(x) &\as c_0\prod_{i=1}^n(x-\alpha_i). \label{eq:p}
	\end{align}
	}
    \bpf
	Recall that the Symbolic Poisson Formula in \refThm{poisson}
	expresses $\res(A,B)$ in terms of the roots $\alpha_i$'s
	of $A$ (modulo $V_a$) and the
	roots $\beta_j$'s of $B$ (modulo $V_b$).
	We now replace $A$ by $\wtp=\sum_{i=0}^n c_ix^{n-i}$,
	and $B$ by $\wtp'= \sum_{i=0}^{n-1} (n-i) c_i x^{n-i-1}$.
	Also, we replace $V_a, V_b$ by $V, V'$ from \refeq{v}.
	Thus
		\begin{align}
			D(\wtp) &= (-1)^{\frac{n(n-1)}{2}}c_0\inv \res(\wtp,\wtp')
				  &\text{(definition of discriminant, see \refeq{D})}
				  \nonumber\\
				&\equiv_1 (-1)^{\frac{n(n-1)}{2}}c_0\inv
				\cdot (nc_0)^n\prod_{i=1}^{n-1}\wtp(\beta_i).
				  &\text{(Poisson Formula, \refThm{poisson}(b),
					modulo $V'$)} \nonumber\\
				&= C\cdot \prod_{i=1}^{n-1}\wtp(\beta_i).
				  &\text{(by definition of $C$, see \refeq{C})}.
				  		\label{eq:Cwtp}
		\end{align}
	Before continuing, we make two observations:
	\bitem
    \item The differentiation $\del[n]$ can be extended
	to the quotient ring $R_0/\bang{V'}_0$
	because $\bang{V'}_0$ is closed\footnote{
	    In contrast, the Viete ideal $V$ for $\wtp(x)$
	    is not closed under $\del[n]$.  }
	under $\del[n]$ (in fact, $\del[n](V')=\set{0}$).
    \item
	$\del[n] \wtp(\beta_i) =$
	\oldNewX{
	$\del[n](\sum_{i=0}^n c_i \beta_i^{n-i})=1$.
	}{
	$\del[n](\sum_{j=0}^n c_j \beta_i^{n-j})=1$.
	}
    \eitem
	\cheeX{Rewrote this derivation:}
	Continuing,
		\begin{align}
		\del[n] D(\wtp) &\equiv_1  C\cdot \sum_{i=1}^{n-1}
				\del[n](\wtp(\beta_i))\cdot\prod_{j\neq i} \wtp(\beta_j)
					& (\text{by differentiation \refeq{Cwtp}})\nonumber\\
			& =  C\cdot \sum_{i=1}^{n-1}
				 \prod_{j\neq i} \wtp(\beta_j)
					& (\text{as observed,}\ \del[n](\wtp(\beta_i))=1)
					\nonumber\\
			& =  C\cdot \sum_{S\in {[n-1]\choose n-2}}
				\prod_{j\in S} \wtp(\beta_j)\nonumber
        \end{align}
        \begin{align}
		\del[n]^2 D(\wtp) &\equiv_1 C\cdot \sum_{S\in{[n-1]\choose n-2}}
				\del[n]\big( \prod_{j\in S} \wtp(\beta_j) \big)
					& (\text{differentiating again})\nonumber\\
			& =  C\cdot \sum_{S\in{[n-1]\choose n-2}}
				\sum_{T\in {S\choose n-3}}\prod_{j\in T}
				\wtp(\beta_j)\nonumber\\
			& =  2!C\cdot \sum_{T\in{[n-1]\choose n-3}}
				\prod_{j\in T} \wtp(\beta_j)
					& \text{(each $T$ arises from two $S$'s)}\nonumber\\
			\vdots&&\nonumber\\
		\del[n]^k D(\wtp) &\equiv_1 (k-1)!C\cdot
				\sum_{S\in{[n-1]\choose n-k}}
				\del[n]\big(\prod_{j\in S} \wtp(\beta_j) \big)
					& (\text{by induction on $k-1$})\nonumber\\
			& =  (k-1)!C\cdot
				\sum_{S\in{[n-1]\choose n-k}}
				\sum_{T\in {S\choose n-1-k}}\prod_{j\in
				T}\wtp(\beta_j)\nonumber\\
			& =  k!C\cdot
				\sum_{T\in{[n-1]\choose n-1-k}}
				\prod_{j\in T} \wtp(\beta_j)
					& \text{(each $T$ arises from $k$ $S$'s)}
				  		\label{eq:delnk}
		\end{align}

	Since $G = \del[n]^{n-m} D(\wtp)$, we conclude
		\beqarrys
		 G &\equiv_1& (n-m)!C \cdot \sum_{S\in{[n-1]\choose m-1}}
						\prod_{j\in S} \wtp(\beta_j)
					&\text{(by \refeq{delnk})}\\
		   &\equiv_2& (n-m)!C \cdot \sum_{S\in{[n-1]\choose m-1}}
						\prod_{j\in S} a_0\prod_{i=1}^n(\beta_j-\alpha_i)
					&\text{(Poisson Formula, \refThm{poisson}(a),
							modulo $\bang{V}_2$)}\\
		   &=& (n-m)!C \cdot \sum_{S\in{[n-1]\choose m-1}}
						\prod_{j\in S} p_\bfalpha(\beta_j)
					&\text{(by definition of $p_\bfalpha$)},
		\eeqarrys
		as claimed by the lemma.
    \epf

	Note that the polynomial $H$ in the Main Theorem
	(see \refeq{Hbfz}) is related to $G$ by a substitution:
		\beql{gh}
			H= \big(c_0^{m+n-2}\big)\inv \cdot
				G\Big|_{c_i\ass(-1)^iz_ic_0~(i=1\dd n)}.\eeql
	This shows that $H\in \ZZ(c_0)[\bfz]$.
	In fact, Lemma B shows that $H\in \ZZ[\bfz]$,
	i.e., $H$ is independent of $c_0$.

	\blemDIY[Lemma B]{\ \\
		The polynomial $H(\bfz)$ is a polynomial
		of degree at most $n+m-2$ in $\ZZ[\bfz]$.
	}

    \bpf
		We know from discriminant theory that
		$D(\wtp)\in \ZZ[\bfc]$
		is homogeneous of total degree $2n-2$ in $c_0\dd c_n$.
		From \refeq{g}
		we conclude that $G=G(c_0\dd c_n)$ is homogeneous of
		degree $(2n-2)-(n-m)=n+m-2$ in $c_0\dd c_n$.
		After the specialization \refeq{c_i},
		$G(c_0,c_1\dd c_n)$ becomes
			\beql{gg}
			G(c_0,-z_1c_0\dd (-1)^iz_i c_0 \dd (-1)^nz_nc_0)
			\eeql
		where each term in \refeq{gg}
		is a product of $c_0^{m+n-2}$ with
		a monomial of degree at most $m+n-2$
		in $\bfz$.  Hence we may divide \refeq{gg} by
		$c_0^{m+n-2}$ to obtain the polynomial $H\in\ZZ[\bfz]$
		(see \refeq{gh}).
    \epf

	\ignore{
    Next consider the following specialization ideal that maps
    the $\alpha_i$'s (roots of $P(x)$) to $\beta_j$'s (roots of $P'(x)$)
    in accordance to the
    multiplicity structure $\bfmu=(\mu_1\dd \mu_n)$:
    let $\beta_1\dd \beta_{m-1}$ be the $m-1$ roots of $P'$
    that are not roots of $P(x)$.

    \begin{tabular}{rlr}
    $W=W(\bfmu)\as\langle$\hspace{-1em}
        &$\alpha_2-\alpha_1\dd \alpha_{\mu_1}-\alpha_1,$
        &$\cdots\cdots(\mu_1)$\\[5pt]
    & $\alpha_{\mu_1+2}-\alpha_{\mu_1+1}\dd
            \alpha_{\mu_1+\mu_2}-\alpha_{\mu_1+1}$,
        &$\cdots\cdots(\mu_2)$\\[5pt]
    & \qquad\qquad$\vdots$ & \\[5pt]
    & $\alpha_{2+\sum_{i=1}^{m-1}
		\mu_i}-\alpha_{1+\sum_{i=1}^{m-1}\mu_i}\dd
            \alpha_{\sum_{i=1}^m\mu_i}-\alpha_{1+\sum_{i=1}^{m-1}\mu_i}$
        &$\cdots\cdots(\mu_m)$\\[5pt]
    & $\beta_{(m-1)+1}-\alpha_1\dd \beta_{(m-1)+\mu_1-1}-\alpha_1,$
        &$\cdots\cdots(\mu_1-1)$\\[5pt]
    & $\beta_{(m+\mu_1-2)+1}-\alpha_{\mu_1+1}\dd
        \beta_{(m+\mu_1-2)+\mu_2-1}-\alpha_{\mu_1+1},$
        &$\cdots\cdots(\mu_2-1)$\\[5pt]
    & \qquad\qquad$\vdots$ & \\[5pt]
    & $\beta_{1+\sum_{i=1}^{m-1}\mu_i}-\alpha_{1+\sum_{i=1}^{m-1}\mu_i}
        \dd \beta_{\mu_m-1+\sum_{i=1}^{m-1}\mu_i}
		-\alpha_{1+\sum_{i=1}^{m-1}\mu_i}
        \rangle$
        & $\cdots\cdots(\mu_m-1)$\\[10pt]
    \end{tabular}
    Let $[W]_2\as \set{[w]_2: w\in W}$ and $R_3\as R_2/\bang{[W]_2}$.
    Again, let $[p]_3$ denote the equivalence class of $[p]_2$
    in $R_3$, and write $p\equiv_3 q$ if $[p]_3=[q]_3$.
    It is obvious that $p\equiv_2 q$ implies $p\equiv_3 q$.
    For the sake of simplicity, let
    \begin{align*}
        r_1=\,&[\alpha_1]_3=\cdots=[\alpha_{\mu_1}]_3\\
           =\,&[\beta_m]_3=\cdots=[\beta_{m+\mu_1-2}]_3,\\
        r_2=\,&[\alpha_{\mu_1+1}]_3=\cdots=[\alpha_{\mu_1+\mu_2}]_3\\
           =\,&[\beta_{(m+\mu_1-2)+1}]_3
		   =\cdots=[\beta_{(m+\mu_1-2)+\mu_2-1}]_3,\\
        &\quad\quad\quad
        \cdots\\
        r_m=\,&[\alpha_{1+\sum_{i=1}^{m-1}\mu_i}]_3
		=\cdots=[\alpha_{\mu_m+\sum_{i=1}^{m-1}\mu_i}]_3\\
           =\,&[\beta_{1+\sum_{i=1}^{m-1}\mu_i}]_3
		   =\cdots=[\beta_{\mu_m-1+\sum_{i=1}^{m-1}\mu_i}]_3.
    \end{align*}
	}

	So far, the symbolic roots $\bfalpha,\bfbeta$ are
	unrelated.  We now show a critical connection
	between them: we introduce
	a polynomial $Q(x)$ defined in terms of the $\beta_j$'s.
	Lemma C shows that, up to $\equiv_3$-equivalence,
	$Q(x)$ can be expressed in terms of the $r_j$'s.
	In other words, we can
	``eliminate the $\beta_i$'s in favor of the $r_j$'s'' in $Q(x)$.

	\blemDIY[Lemma C]{
		Let $Q(x)\as \prod_{j=1}^{m-1}(x-\beta_j)$.
		Then
			$$Q(x)\equiv_3
			\efrac{n}\sum_{k=1}^m \mu_k
				\prod_{\substack{j=1\\j\neq k}}^m \Big(x-r_j\Big).
			$$
		In particular,
			\beql{c}
			Q(r_i) = \frac{\mu_i}{n}
				\prod_{\substack{j=1\\j\neq i}}^m (r_i-r_j).
				\eeql
	}

	\bpf
	   \begin{align}
		   \wtp'(x)&\equiv_2~ nc_0\prod_{j=1}^{n-1}(x-\beta_j)
		   				\notag\\
		   	&\equiv_3~ nc_0\prod_{j=1}^{m-1}(x-\beta_j)
						\cdot \prod_{k=1}^{m}(x-r_k)^{\mu_k-1}
						\notag\\
	\text{Hence\hspace*{1.1in}}\notag\\
		   Q(x)&\equiv_3~
		   			\frac{\wtp'(x)}{
						nc_0\prod_{k=1}^{m}(x-r_k)^{\mu_k-1}}
					\label{eq:q(x)}\\
	\text{On the other hand,\hspace*{0.5in}}\notag\\
			\wtp'(x)=(\wtp(x))' &\equiv_3~c_0\Big(\prod_{k=1}^m (x-r_k)^{\mu_k}\Big)'
						\notag\\
				&=~ c_0\sum_{k=1}^m
						\mu_k(x-r_k)^{\mu_k-1}
					\Big(\prod_{\substack{j=1\\ j\neq k}}^m
						(x-r_j)^{\mu_j}\Big)
						\notag\\
	\text{Plugging this into \refeq{q(x)}\hspace*{0.2in}}\notag\\
		   Q(x)&\equiv_3~
				\efrac{n}\sum_{k=1}^m \mu_k
						\prod_{\substack{j=1\\j\neq k}}^m (x-r_j)
					\label{eq:q(x)k}
		\end{align}
	This gives our ``elimination result'' on $Q(x)$.
	Furthermore, when we evaluate $Q(x)$ at $x=r_i$,
	the $k$th summand in \refeq{q(x)k} vanishes unless $k=i$.
	This yields
		   $Q(r_i) \equiv_3 \tfrac{\mu_i}{n}
					\prod_{\substack{j=1\\j\neq i}}^m (r_i-r_j)$.
	\epf

	In our final Lemma D, we exploit the ``$\beta$-elimination''
	property of Lemma C to prove that
	$H(\bfz)$ is the $\bfmu$-gist of $\dplus_\bfmu(\bfr)$.

	\blemDIY[Lemma D]{
		The polynomial $H(\bfz)$ of Lemma B has the property
			$$H(\ole_1\dd\ole_n)\equiv_3
				C_\bfmu\cdot\dplus_\bfmu(\bfr).$$
		In other words, $(C_\bfmu)\inv H(\bfz)$
		is the $\bfmu$-gist of $\dplus_\bfmu(\bfr)$.
	}

    \bpf
	From the definition of $R_3=R_2/\bang{W\cup W'}$ in \refeq{r3},
	we see that the polynomial $\wtp$ has exactly $m$ distinct roots,
	namely, $[r_1]_3\dd [r_m]_3$ in $R_3$.
	Among the roots $\bfbeta$ of $\wtp'$,
	exactly $m-1$ roots are different from the $[r_j]$'s,
	namely, $[\beta_1]_3\dd [\beta_{m-1}]_3$.
	From \refeq{p},
		$$p_\bfalpha(\beta_j)=c_0\prod_{i=1}^n(\beta_j-\alpha_i)
				\equiv_3 c_0\prod_{i=1}^m (\beta_j-r_i)^{\mu_i}.$$
	Moreover,
    	$p_\bfalpha(\beta_j)\equiv_3 0$ iff $j\geq m$.
	It follows that for $S\in {[n-1]\choose m-1}$,
			\beql{g3}
				\prod_{j\in S} p_\bfalpha(\beta_j)
				\equiv_3 \clauses{ 0 & \rmif\ S\neq\set{1\dd m-1}\\
					\prod_{j=1}^{m-1} p_\bfalpha(\beta_j) & \ELSe.}
			\eeql
	Hence
		\begin{align}
			G &\equiv_2 (n-m)!C\cdot  \sum_{S\in{[n-1]\choose m-1}}
				\prod_{j\in S} p_\bfalpha(\beta_j)
						& \text{(Lemma A)} \notag\\
				&\equiv_3 (n-m)!C\cdot
					\prod_{j=1}^{m-1} p_\bfalpha(\beta_j).
			& \text{(by \refeq{g3})} &\label{eq:g3a}\\
        H(\ole_1\dd\ole_n)
        	&=~\efrac{c_0^{m+n-2}}G(c_0,-\ole_1c_0\dd (-1)^n\ole_nc_0)
				&\text{(by definition of $H(\bfz)$)} \notag\\
			&\equiv_3~\frac{(n-m)!C\cdot
                    \prod_{j=1}^{m-1}p_{\bfalpha}(\beta_j)
                    }{c_0^{m+n-2}}
					& \text{(by \refeq{g3a})} \notag\\
			&\equiv_3~\frac{(n-m)!C\cdot
                c_0^{m-1}\prod_{j=1}^{m-1}
					\prod_{i=1}^m (\beta_j-r_i)^{\mu_i}}
            	{c_0^{m+n-2}}\notag\\
			&=~(n-m)!(-1)^{n\choose 2} n^n\cdot
                \prod_{j=1}^{m-1}
					\prod_{i=1}^m (\beta_j-r_i)^{\mu_i}
				& \text{(by expanding $C$)} \notag
        \end{align}
        \begin{align}
			&=~(n-m)!(-1)^{n(m-1)+{n\choose 2}}n^n\cdot
            	\prod_{i=1}^{m}\prod_{j=1}^{m-1}
            		(r_i-\beta_j)^{\mu_i}
				&\text{(negate $(\beta_j-r_i)$)} \notag\\
			&=~(n-m)!(-1)^{n(m-1)+{n\choose 2}}\cdot n^n\cdot
				\prod_{i=1}^{m}\big(Q(r_i)\big)^{\mu_i}
				&\text{(by definition of $Q(x)$)}
					\label{eq:evalHri} \\
	n^n\cdot \prod_{i=1}^{m}\big(Q(r_i)\big)^{\mu_i}
			&\equiv_3~ n^n\cdot\prod_{i=1}^{m}\Big(\frac{\mu_i}{n}
					\prod_{\substack{k=1\\k\neq i}}^{m}(r_i-r_k)
							\Big)^{\mu_i}
					&\text{(by Lemma C)}\notag\\
			&=~ n^n\cdot\Big(\prod_{i=1}^{m}
				(\tfrac{\mu_i}{n})^{\mu_i}\Big)\cdot
					\prod_{i=1}^{m}\Big(
				\prod_{\substack{k=1\\k\neq i}}^{m}
					(r_i-r_k)\Big)^{\mu_i} \notag\\
			&=~ \Big(\prod_{i=1}^{m}\mu_i^{\mu_i}\Big)\cdot
				\prod_{1\leq i<k\leq m}(r_i-r_k)^{\mu_i}
						(r_k-r_i)^{\mu_k}
					&\text{(since $n=\mu_1+\cdots +\mu_m$)}\notag
        \end{align}
        \begin{align}
			\hspace*{1.1in}&=~ \Big(\prod_{i=1}^{m}\mu_i^{\mu_i}\Big)\cdot
				\prod_{1\leq i<k\leq m}(r_i-r_k)^{\mu_i+\mu_k}
						(-1)^{\mu_k} \notag\\
			&=~ \Big(\prod_{i=1}^{m}\mu_i^{\mu_i}\Big)\cdot
				(-1)^{\sum_{k=1}^m(k-1)\mu_k}\cdot
            	\prod_{1\leq i<k\leq m}(r_i-r_k)^{\mu_i+\mu_k}
				\notag\\
			 &\text{\hspace*{1in}
			 	(since each $(-1)^{\mu_k}$ appears $k-1$ times)}
			 	\notag\\
			&=~
				\Big(\prod_{i=1}^{m}\mu_i^{\mu_i}\Big)
				(-1)^{\sum_{k=1}^m(k-1)\mu_k}\cdot
					\dplus_\bfmu(\bfr).
			 &\text{(by definition of $\dplus_\bfmu$)} \notag
        \end{align}
        In summary,
        \begin{align*}
			H(\ole_1\dd\ole_n)
				&\equiv_3~ C_\bfmu \dplus_\bfmu(\bfr)
        \end{align*}
		where $C_\bfmu =(n-m)! (-1)^N
						\prod_{i=1}^{m}\mu_i^{\mu_i}$ and
        \begin{align*}
			N &=~ n(m-1)+{n\choose 2}+\sum_{k=1}^m (k-1)\mu_k \\
			  &\equiv~ nm+{n\choose 2}+\sum_{k=1}^m k\mu_k.
        \end{align*}
	\cheeX{Best to avoid "mod 2", which is obvious.
		But I would love to simplify this expression if I can!
		}
		We conclude from \refPro{symmetric}
		that $(C_\bfmu)\inv H(z_1\dd z_n)$ is the
		$\bfmu$-gist of $\dplus_\bfmu(\bfr)$.
    \epf

	With Lemma D, we conclude the proof of the Main Result.
	The Main Result yields two corollaries:

	\bgenDIY{\sc First Corollary}{
    		\label{cor:denominator}
			\ \\
			Let $p(x)=\sum_{i=0}^n a_ix^{n-i}\in\ZZ[x]$
			with multiplicity vector $\bfmu=(\mu_1\dd \mu_m)$.
			Then $\dplus(p)$ is a rational number whose
			denominator is at most
				\beql{dplusp}
					(n-m)!\big(\prod_{i=1}^m\mu_i^{\mu_i}\big)
					a_0^{n+m-2}.
				\eeql
			This bound is sharp.
		}

	\bpf
	From Lemma D,
		$$\dplus(p) = (C_\bfmu)\inv H(\ole_1\dd\ole_n)$$
	where $\ole_i = (-1)^i a_i/a_0$.
	By Lemma B, $H(\bfz)$ is an integer polynomial of degree
	$\le n+m-2$ in $\bfz$.  Hence $H(\ole_1\dd \ole_n)$
	is a rational number with denominator at most $a_0^{n+m-2}$.
	Thus
		$\dplus(p) = \frac{1}{C_{\bfmu}}
						\cdot H(\ole_1\dd\ole_n)
					\ge \frac{1}{|C_{\bfmu}|} a_0^{-n-m+2}$.
	Our lemma follows since
	$|C_\bfmu|=(n-m)!\prod_{i=1}^m\mu_i^{\mu_i}$.

	To see the sharpness of this denominator, note that the
	denominator is exactly $|C_\bfmu| \cdot a_0^{n+m-2}$
	unless there is cancellation with the numerator.
	Moreover, the value $C_\bfmu$ as well as
	the degree $n+m-2$ of $H(\bfz)$ are exact.
	\epf

	\bgenDIY{\sc Second Corollary}{
    		\label{cor:gistequal}
		\ \\
		Let $\bfmu_1, \bfmu_2$ be two $m$-partitions of $n$.
		\\
		Then $\dplus_{\bfmu_1}(\bfr)$ and $\dplus_{\bfmu_1}(\bfr)$
	    differ by a constant multiplier. More precisely,
			$$C_{\bfmu_1} \dplus_{\bfmu_1}(\bfr)
				= H_{n,m}(\ole_1\dd\ole_n)
				=C_{\bfmu_2}\dplus_{\bfmu_2}(\bfr).$$
		}

\sect{Application to the Complexity of Root Clustering}
\label{sec:complexitybound}
	As described in the introduction, the bit-complexity
	of the root clustering
	algorithm in Becker et al.~\cite{becker+4:cluster:16}
	in the ``benchmark case'' has an additive term of the form
	$n \log \big(|\dplus(p)|\inv\big)$
	where $n=\deg(p)$.
	This quantity is bounded by
	\refCor{denominator}
	in terms of $n, \bfmu(p)$ and $a_0$,
	the leading coefficient of $p$.
	\ignore{
	    No such lower bound is possible if $p(x)$ has arbitrary
	    complex coefficients.
	    {\em Now suppose $p(x)$ is an integer polynomial.}
	    Of course, $\wt{p}$ may not be an integer polynomial,
	    but it is (theoretically) possible to obtain a lower
	    bound on $|\dplus(\wt{p})|$.  Such a bound
	    would be practically worthless because it
	    would be extremely small.
	    Here is an indirect argument to get a reasonable lower bound.
		Let $\sigma(p)$ be the minimum separation between distinct roots
		of $p$.  If $\vareps \le \sigma(p)$, then it is clear that
	    there is only one root in each cluster and so $\wt{p}=p$.
	    The lower bound for $\dplus(p)$ is now viable.
	    What if $\vareps>\sigma(p)$?  In this case, the complexity of
	    the algorithm is still no worse than the case
	    where $\vareps\le \sigma(p)$.  This follows
	    from the nature of our algorithm.
	}
	Since $\bfmu(p)$ is not easily deduced from the input $p$,
	our propose to bound the quantity
			$$F_{n,m} \as \max_{\bfmu} \prod_{i=1}^m\mu_i^{\mu_i}$$
	where $\bfmu$ range over all partitions of $n$
	with exactly $m$ parts.
	We consider a closely related maximization problem
		\begin{align}
			\Phi_{n,m} &\as \max_{\bfx \in \Omega_{n,m}}
							\Phi(\bfx)\\
			\text{where}\hspace*{0.5in}& \notag\\
				\vphi(x) &\as x\log x, \notag\\
			\Phi(\bfx) &= \Phi(x_1\dd x_m) \as \sum_{i=1}^m \vphi(x_i),\\
			\Omega_{n,m} &\as \set{\bfx \in\RR^m:
			n-m+1\ge x_1\ge x_2\ge\cdots\ge x_m\ge 1}.
		\end{align}
	Let $\bfx^*\in \Omega_{n,m}$ denote any maximum point of $\Phi(\bfx)$,
	i.e., $\Phi(\bfx^*)=\Phi_{n,m}$.
	Since
	$\Phi_{n,m}$ is a maximization over the real set $\Omega_{n,m}$
	while $F_{n,m}$ is a maximization over integer points
	$(\mu_1\dd \mu_m)$ of $\Omega_{n,m}$, we have the inequality
		\beql{logF}
			\log (F_{n,m})\le \Phi_{n,m}.\eeql
	Here and following, assume $\log$ is the
	natural logarithm $\log=\ln$.
	In fact, we do have equality:

    \bthml{boundary}
		The value $\Phi_{n,m}$ is attained at the
		unique point $\bfx^*=(n-m+1,1\dd 1)\in\Omega_{n,m}$:
			$$\Phi_{n,m} = \Phi(\bfx^*)
				 =(n-m+1)\log(n-m+1).$$
		Hence $\log(F_{n,m}) = \Phi_{n,m}$.
    \ethml
	\bpf
	We first prove that the function $\Phi(\bfx)$
	achieves its maximum on the boundary of $\Omega_{n,m}$.
	Let $\del[i]\Phi = 1+\log x_i$ be
	the partial derivative with respect to $x_i$
	($i=1\dd n$).  The critical point $\bfx_0$ of
	$\Phi(\bfx)$ is where $\del[i]\Phi(\bfx_0)=0$ for all $i=1\dd m$.
	So $\bfx_0=(e\inv\dd e\inv)$ where $e\inv=0.3679$.
	The second partial derivatives
	are given by $\del[ij]\Phi = \delta_{ij} \efrac{x_i}$
	where $\delta_{ij}$ is Kronecker's delta.
	So the Hessian matrix is $H(\bfx)=(\del[ij]\Phi)_{i,j=1}^n$
	is the diagonal matrix $\diag(x_1\inv\dd x_n\inv)$.
	Clearly $H(\bfx_0)$ is a positive definite matrix.
	This proves that
	$\bfx_0$ is a minimum point.  Since there are no other critical
	points, we conclude that $\Phi(\bfx)$ attains its maximum value
	in $\Omega_{n,m}$ at some boundary point.

	Let $\bfx^*=(x^*_1\dd x^*_m)\in \Omega_{n,m}$ be any maximum point.
	If $x^*_m>1$ then $n-m+1>x^*_1\ge x^*_2\ge\cdots\ge x^*_m>1$,
	and so $\bfx^*$ lies in the interior
	of $\Omega_{n,m}$, contradiction.  Therefore $x^*_m=1$.
	Let $\bfx'=(x^*_1\dd x^*_{m-1})$,
	and observe that $\bfx'\in\Omega_{n-1,m-1}$.
	Moreover, $\Phi_{n,m} = \sum_{i=1}^m \vphi(x^*_i)
				=\sum_{j=1}^{m-1} \vphi(x^*_j)$.
	Consider the injective map
		$$\iota_m: \Omega_{n-1,m-1}\to \Omega_{n,m}$$
	where $\iota_m(y_1\dd y_{m-1})=(y_1\dd y_{m-1},1)$.
		\begin{align*}
			\Phi_{n,m} &= \Phi(\bfx^*)\\
					&= \max_{\bfx\in\iota_m(\Omega_{n-1,m-1})} \Phi(\bfx)
			 	&\text{(since $\bfx^*\in\iota_m(\Omega_{n-1,m-1})$)}\\
					&= \max_{\bfy\in\Omega_{n-1,m-1}} \Phi(\bfy)\\
					&= \Phi_{n-1,m-1}.
		\end{align*}
		By induction on $k=1\dd m-1$,
		we conclude that $\Phi_{n,m}=\Phi_{n-k,m-k}$.
		When $k=m-1$, we see that $\Phi_{n-m+1,1}=\vphi(n-m+1)$
		as claimed.  The maximum point $x^*$ in $\Omega_{n-m+1,1}\ib\RR$
		is unique: $x^*=n-m+1$.  By taking the inverses
		of the $\iota_k$'s, we conclude that the maximum point
		in $\Omega_{n,m}$ is $\bfx^*=(n-m+1,1\dd 1)$.
	\epf

    \bcorl{logd}
	\ \\ Let $p$ be an integer polynomial of degree $n$ with
	an $L$-bit leading coefficient $a_0$. Then
		$$\log(|\dplus(p)|\inv)
			\le 2n(\log n + L\log(2)) = \wtO(nL)$$
	\ecorl
	\bpf
		\begin{align*}
			\log(|\dplus(p)|\inv)  &\le
				\log\big((n-m)!\big(\prod_{i=1}^m\mu_i^{\mu_i}\big)
					a_0^{n+m-2}\big)
						& \text{(from \refeq{dplusp})}\\
				&\le \log((n-m)!) + \vphi(n-m+1) + (n+m-2)\log(2^L)
					& \text{(from \refThm{boundary})}\\
				& \le (n\log n)+ (n\log n) + (2n)L\log(2).
		\end{align*}
	\epf

	\ignore{%
		We first prove an auxiliary result:
	
	    \bleml{kfunction}
	    Let $f(x)\as (n-x)\ln\frac{n-x}{m-x}$
		where $n\geq m$. Then
	    $f'(x)\geq 0$ for $x\in[0,m)$.
	    \eleml
	
	    \bpf
	    Note that
	    $f'(x)=-\ln\frac{n-x}{m-x}+\frac{n-m}{m-x}$
	    can be rewritten as
	    	$f'(x)= \frac{n-m}{m-x}-\ln\Big(1+\frac{n-m}{m-x}\Big)$.
		Let $g(t)=t-\ln(1+t)$ and $t(x)=\frac{n-m}{m-x}$.
	    Then $f'(x)=g(t(x))$.
	    Since $g(t)\geq0$ when $t\geq0$
	    and $\frac{n-m}{m-x}\geq0$ for $x\in[0,m)$,
	    $f'(x)\geq0$ (where equality holds iff $n=m$).
	    \epf
	
	    Now we are ready to prove \refThm{boundary}.
	
	    \bpf
	    Note that maximizing $F(\bfx)$ subjected to
	    $x_i\geq 1~(1\leq i \leq m)\wedge \sum_{i=1}^mx_i=n$
	    is equivalent to maximizing $\ln F(\bfx)$
	    under the same constraints,
	    which is also equivalent to the following problem:
	    \begin{align}
	    	\max\quad &f(\bfx)=\sum_{i=1}^{m-1}x_i\ln{x_i}
	    		+(n-\sum_{i=1}^{m-1}x_i)\ln(n-\sum_{i=1}^{m-1}x_i)
					\label{eq:optimization}\\
	    	\mbox{subject to}\quad
	    		&x_i\geq1~(1\leq i \leq m-1)
					\label{eq:constraint1}\\
	    		&\sum_{i=1}^{m-1}x_i\leq n-1.
					\label{eq:constraint2}
	    \end{align}
	    Thus proving $\max F(\bfx)=(n-m+1)^{n-m+1}$
	    is equivalent to proving
	    $\max f(\bfx)=(n-m+1)\ln(n-m+1)$.
		Moreover, observing that the maximum value of
		\eqref{eq:optimization}
	    is a function in $n$ and $m$,
	    we introduce $M(n,m)\as \max f(\bfx)$ satisfying Constraints
	    \eqref{eq:constraint1} and \eqref{eq:constraint2}.
	
		The proof of $M(n,m)=(n-m+1)\ln(n-m+1)$ can be divided into three
		steps.
	    \benum
	    \item We show that $M(n, m)=\max(n\ln\frac{n}{m},M(n-1,m-1))$.
	
	    When optimizing $f(\bfx)$, we first take the partial derivatives
	    of $f(\bfx)$ in terms of $x_i$'s and set up the following system
	    to compute the critical points
	    \[\frac{\partial f}{\partial x_i}=
	        \ln\frac{x_i}{n-\sum_{j=1}^{m-1}x_j}, i=1\dd m-1.\]
	    This system has only one critical point which is
	    $x_1=\cdots=x_{m-1}=n/m$ and $f(n/m\dd n/m)=n\ln\frac{n}{m}$.
	    Next we will compare $f(n/m\dd n/m)$ with the maximum value of
		$f(\bfx)$ on boundaries to determine
	    which one is the optimal value we are looking for.
	
	    Without loss of generality, we assume that
	    the maximum value on boundaries will either be achieved
	    when $x_{m-1}=1$ or $n-\sum_{i=1}^{m-1}x_i=1$.
	    In the first case, we are going to search for
	    the solution of the following problem
	    \begin{align*}
	    \min\quad &\sum_{i=1}^{m-2}x_i\ln{x_i}
	    +(n-1-\sum_{i=1}^{m-2}x_i)\ln(n-1-\sum_{i=1}^{m-2}x_i)\\
	    \mbox{subject to}\quad
	    &x_i\geq1~(1\leq i \leq m-2)\\
	    &\sum_{i=1}^{m-1}x_i\leq n-2.
	    \end{align*}
	    In the second case, the problem we are concerned with is
	    \begin{align*}
	    \max\quad &\sum_{i=1}^{m-1}x_i\ln{x_i}\\
	    \mbox{subject to}\quad
	    &x_i\geq1~(1\leq i \leq m-1)\\
	    &\sum_{i=1}^{m-1}x_i= n-1,
	    \end{align*}
	    which is equivalent to
	    \begin{align*}
	    \max\quad &\sum_{i=1}^{m-2}x_i\ln{x_i}
	    +(n-1-\sum_{i=1}^{m-2}x_i)\ln(n-1-\sum_{i=1}^{m-2}x_i)\\
	    \mbox{subject to}\quad
	    &x_i\geq1~(1\leq i \leq m-1)\\
	    &\sum_{i=1}^{m-1}x_i\leq n-2.
	    \end{align*}
	    For both cases, the maximum value is exactly $M(n-1,m-1)$.
	    Therefore,
	    \[M(n,m)=\max\big(n\ln\frac{n}{m}, M(n-1,m-1)\big).\]
	    \item We show that $M(n-k,m-k)=(n-m+1)\ln(n-m+1)$ by deduction
	    on $k$.
	
	    When $k=m-1$, $M(n-k,m-k)=M(n-m+1,1)$. Since the only point
	    in the feasible set is$\bfx=(x_1)=(n-m+1)$, the maximum value
	    is
	        \[M(n-m+1,1)=(n-m+1)\ln(n-m+1).\]
	
	    Assume $M(n-\ell,m-\ell)=(n-m+1)\ln(n-m+1)$ holds for
		$m-1\geq\ell>k\geq 0$.
	    By \refLem{kfunction},
	    $f(x)=(n-x)\ln\frac{n-x}{m-x}$ is monotonically increasing
	    over $[0, m-1]$. Therefore,
	    $(n-k)\ln\frac{n-k}{m-k}\leq [n-(m-1)]\ln\frac{n-(m-1)}{m-(m-1)}
	    =M\big(n-(k+1),m-(k+1)\big)$.
	    It follows that
	    \begin{align*}
	    M(n-k,m-k)&=\max\Big((n-k)\ln\frac{n-k}{m-k},
			M\big(n-(k+1),m-(k+1)\big)\Big)\\
	            &=M\big(n-(k+1),m-(k+1)\big)\\
	            &=(n-m+1)\ln(n-m+1).
	    \end{align*}
	    \item Now we immediately conclude that
	    the optimal value for \eqref{eq:optimization}
		under the constraints \eqref{eq:constraint1} and
		\eqref{eq:constraint2}
	    is $M(n,m)=(n-m+1)\ln(n-m+1)$. Therefore,
	    the optimal value for $F(\bfx)$ is $(n-m+1)^{n-m+1}$,
	    which is achieved at $x_1=\cdots=x_{m-1}=1$ and $x_m=n-m+1$.
	    \eenum
	    \epf
	}%
    \ignore{%
	        \bthml{boundary}
	    Let $\bfx=(x_1\dd x_m)$
	    and $F(\bfx)=\prod_{i=1}^mx_i^{x_i}$.
	    Then the optimal solution to the following problem
	    \[\min F(\bfx)\quad\mbox{s.t.}\quad
	    	x_i\geq 1~(1\leq i \leq m)
				\wedge \sum_{i=1}^mx_i=n~\mbox{where}~m\leq n\]
	    is $x_1=\cdots=x_m=n/m$.
	    Moreover, the minimum value of $F(\bfx)$ is
	    $\Big(\frac{n}{m}\Big)^n$.
	    \ethml
	
	    For proving the theorem, we present the following lemma first.
	
	    \bleml{kfunction}
	    If $n\geq m$,
		$(n-k)\ln\frac{n-k}{m-k}\geq[n-(k+1)]\ln\frac{n-(k+1)}{m-(k+1)}$
	    for $k=0\dd m-1$ and $n>1$.
	    \eleml
	
	    \bpf
	    Let $f(x)=(n-x)\ln\frac{n-x}{m-x}$ where $0\leq x\leq m-1$.
	    Take the derivative of $f(x)$ for $x$ and we get
	    $f'(x)=-\ln\frac{n-x}{m-x}+\frac{n-m}{m-x}$.
	    It can be rewritten to
	    $f'(x)=
	    \frac{n-m}{m-x}-\ln\Big(1+\frac{n-m}{m-x}\Big)$.
	    Since $\frac{n-m}{m-x}>0$ for $x\in[1,m-1]$,
	    $f'(x)\geq0$ (``$=$" holds only when $n=m$).
	    Hence
	    $(n-k)\ln\frac{n-k}{m-k}
	        \geq[n-(k+1)]\ln\frac{n-(k+1)}{m-(k+1)}$.
	
	    \epf
	
	    Now we are ready to prove \refThm{boundary}.
	
	    \bpf
	    Note that minimizing $F(\bfx)$ subjected to
	    $x_i\geq 1~(1\leq i \leq m)\wedge \sum_{i=1}^mx_i=n$
	    is equivalent to maximizing $\ln F(\bfx)$
	    under the same constraints,
	    which is also equivalent to the following problem:
	    \begin{align}
		    \min\quad &f(\bfx)=\prod_{i=1}^{m-1}x_i\ln{x_i}
		    	+(n-\sum_{i=1}^{m-1}x_i)\ln(n-\sum_{i=1}^{m-1}x_i)
				\label{eq:optimization}\\
		    \mbox{subject to}\quad
		    &x_i\geq 1~(1\leq i \leq m-1)\label{eq:constraint1}\\
		    &\sum_{i=1}^{m-1}x_i\leq n-1.\label{eq:constraint2}
	    \end{align}
	    Thus proving $\min F(\bfx)=\big(\frac{n}{m}\big)^n$
	    is equivalent to proving
	    $\min f(\bfx)=n\ln\frac{n}{m}$.
	    Moreover, observing that \eqref{eq:optimization} only
		depends on $n$ and $m$,
	    we introduce $M(n,m)\as \min f(\bfx)$ satisfying Constraints
	    \eqref{eq:constraint1} and \eqref{eq:constraint2}.
	    We first establish the relationship between $M(n, m)$
	    and $M(n-1,m-1)$.
	
	    When optimizing $f(\bfx)$, we first take the partial derivatives
	    of $f(\bfx)$ in terms of $x_i$'s and set up the following system
	    to compute the critical points
	    \[\frac{\partial f}{\partial x_i}=
	        \ln\frac{x_i}{n-\sum_{i=1}^{m-1}x_i}, i=1\dd m-1.\]
	    This system has only one critical point which is
	    $x_1=\cdots=x_{m-1}=n/m$ and $f(n/m\dd n/m)=n\ln\frac{n}{m}$.
	    Next we will compare $f(n/m\dd n/m)$ with the minimum value of
		$f(\bfx)$ on boundaries to determine
	    which one is the optimal value we are looking for.
	
	    Without loss of generality, we assume that
	    the minimum value on boundaries will either be achieved
	    when $x_{m-1}=1$ or $n-\sum_{i=1}^{m-1}x_i=1$.
	    In the first case, we are going to search for
	    the solution of the following problem
	    \begin{align*}
	    \min\quad &\prod_{i=1}^{m-2}x_i\ln{x_i}
	    +(n-1-\sum_{i=1}^{m-2}x_i)\ln(n-1-\sum_{i=1}^{m-2}x_i)\\
	    \mbox{subject to}\quad
	    &x_i\geq 1~(1\leq i \leq m-2)\\
	    &\sum_{i=1}^{m-1}x_i\leq n-2.
	    \end{align*}
	    In the second case, the problem we are concerned is
	    \begin{align*}
	    \min\quad &\prod_{i=1}^{m-1}x_i\ln{x_i}\\
	    \mbox{subject to}\quad
	    &x_i\geq 1~(1\leq i \leq m-1)\\
	    &\sum_{i=1}^{m-1}x_i= n-1,
	    \end{align*}
	    which is equivalent to
	    \begin{align*}
	    \min\quad &\prod_{i=1}^{m-2}x_i\ln{x_i}
	    +(n-1-\sum_{i=1}^{m-2}x_i)\ln(n-1-\sum_{i=1}^{m-2}x_i)\\
	    \mbox{subject to}\quad
	    &x_i\geq 1~(1\leq i \leq m-1)\\
	    &\sum_{i=1}^{m-1}x_i\leq n-2.
	    \end{align*}
	    For both cases, the minimum value is exactly $M(n-1,m-1)$.
	    Therefore,
	    \[M(n,m)=\min\big(n\ln\frac{n}{m}, M(n-1,m-1)\big).\]
		Next we prove that $M(n-k,m-k)=(n-k)\ln\frac{n-k}{m-k}$
		by deduction on $k$.
	
	    When $k=m-1$, $M(n-k,m-k)=M(n-m+1,1)$. Since the only point
	    in the feasible set is $x_1=n-m+1$, the minimum value
	    is $M(n-m+1,1)=(n-m+1)\ln(n-m+1)$.
	    Assume $M(n-\ell,m-\ell)=(n-\ell)\ln\frac{n-\ell}{m-\ell}$ holds for $\ell>k$.
	    By \refLem{},
	    $(n-k)\ln\frac{n-k}{m-k}\geq [n-(k+1)]\ln\frac{n-(k+1)}{m-(k+1)}
	    =M(n-(k+1),m-(k+1))$.
	    Then
	    \[M(n-k,m-k)=(n-k)\ln\frac{n-k}{m-k}.\]
	
	    Now we immediately conclude that
	    the optimal value for \eqref{eq:optimization}
	    under the constraints \eqref{eq:constraint1} and \eqref{eq:constraint2}
	    is $M(n,m)=n\ln\frac{n}{m}$. Therefore,
	    the optimal value for $F(\bfx)$ is $\Big(\frac{n}{m}\Big)^n$,
	    which is achieved at $x_1=\cdots=x_m=n/m$.
	    \epf
	
	    By \refThm{boundary}, we immediately get the following result.
	
	    \bcor
	    Let $\bfmu=(\mu_1\dd\mu_m)$ be a partition of $n$.
	    $\prod_{i=1}^m\mu_i^{\mu_i}\geq\Big(\frac{n}{m}\Big)^n$
	    and ``$=$" holds iff $m\mid n$ and $\mu_1=\cdots=\mu_m=n/m$.
	    \ecor
    }%
\ignore{
\sssect{Determine the number of distinct roots}
    \bthm
    Given $P(x)=\sum_{i=0}^nc_{n-i}x^i\in K[x]$, then $P$ has exactly
    $m$ distinct roots iff
    $\frac{\partial^i D}{\partial c_n^i}=0$
    for $i=0\dd n-m-1$ and
    $\frac{\partial^{n-m} D}{\partial c_n^{n-m}}\neq0$
    where
    $D$ is the Sylvester discriminant of $P(x)$.
    \ethm

    \bpf
    ($\Rightarrow$): Suppose $P(x)$ has exactly $m$ roots,
    say $r_1\dd r_m$, and the multiplicity structure is
    $\bfmu=(\mu_1\dd\mu_m)$ respectively.
    According to \refThm{dplusismusymmetric},
    $\frac{\partial^{n-m} D}{\partial c_n^{n-m}}
    |_{c_i=(-1)^i\ole_ic_0}=cc_0^{m+n-2}\dplus(\bfmu)\neq 0$
    where $c$ is a constant determined by $\bfmu$.
    Thus $\frac{\partial^{n-m} D}{\partial c_n^{n-m}}\neq0$.

    From the proof of \refThm{dplusismusymmetric},
    $\frac{\partial^i D}{\partial c_n^i}=0$
    for $i=0\dd n-m-1$.

    ($\Leftarrow$) If $P(x)$ has $m'$ distinct roots with $m'>m$,
    then $n-m'<n-m$. According to the above deduction,
    $\frac{\partial^{n-m'} D}{\partial c_n^{n-m'}}\neq0$,
    which contradicts with the assumption that
    $\frac{\partial^i D}{\partial c_n^i}=0$
    for $i=0\dd n-m-1$.

    If $P(x)$ has $m'$ distinct roots with $m'< m$, then
    $\frac{\partial^{n-m'} D}{\partial c_n^{n-m'}}\neq0$.
    Since $n-m<n-m'$,
    $\frac{\partial^{n-m} D}{\partial c_n^{n-m}}=0$,
    which leads to a contradiction.
    \epf
    }%
\sect{Final Remarks}\label{sec:conclusion}
    \revisedX{
    The Main Result of this paper proves the conjecture that the
	root function $\dplus_\bfmu$ is $\bfmu$-symmetric,
	by giving a determinantal formula for its gist.
	This implies an exact formula for the
	D-plus discriminant $\dplus(p)$ of a polynomial $p(x)$
	in terms of the coefficients of $p(x)$.
	This leads to an exact lower bound on $|\dplus(p)|$
	when $p$ is an integer polynomial.
	Our Main Result is established using an ideal-theoretic
	proof.  For this, we need to re-cast the classic Poisson
	formula as the ``symbolic Poisson formula'',
	which has independent interest.
	Finally, we apply our result to give a very sharp
	bound on the term
	$n\log\big(|\dplus(p)|\inv\big)$
	arising in the bit complexity of the root clustering algorithm of
	Becker et al.~\cite{becker+4:cluster:16}.
    }

\section*{Acknowledgments}
The authors would like to thank Professor Dongming Wang for his
generous support and the use of the resources of the SMS International.
This work is done during Chee's sabbatical year in China with
the generous support of GXUN, Beihang University and Chinese Academy
of Sciences in Beijing.

\bibliographystyle{siamplain}

\end{document}